\documentclass[aps,preprintnumbers,nofootinbib,superscriptaddress,twocolumn]{revtex4}

\usepackage{graphicx}
\usepackage{epsfig,graphics}
\usepackage{amsfonts,amssymb,amsmath}
\usepackage{amsthm}

\newcommand{\comment}[1]{}
\newcommand{\ket}[1]{\left |  #1 \right\rangle}
\newcommand{\bra}[1]{\left \langle #1  \right |}

\newcommand{\ketbra}[2]{|#1\rangle\!\langle#2|}
\newcommand{\id}{\openone}
\newcommand{\ot}{\otimes}
\newcommand{\cF}{\mathcal{F}}
\newcommand{\hP}{\hat{P}}
\newcommand{\hQ}{\hat{Q}}
\newcommand{\hp}{\hat{p}}
\newcommand{\hq}{\hat{q}}
\theoremstyle{plain}
\newtheorem{theorem}{Theorem}

\theoremstyle{definition}
\newtheorem{definition}{Definition}

\begin{document}

\title{Private Randomness Expansion With Untrusted Devices}

\author{Roger \surname{Colbeck}}
\email[]{rcolbeck@perimeterinstitute.ca}
\affiliation{Perimeter Institute for Theoretical Physics, 31 Caroline Street North, Waterloo, ON N2L 2Y5, Canada.}

\author{Adrian \surname{Kent}}
\email[]{a.p.a.kent@damtp.cam.ac.uk}
\affiliation{Centre for Quantum Information and Foundations, DAMTP, Centre for
  Mathematical Sciences, University of Cambridge, Wilberforce Road,
  Cambridge, CB3 0WA, U.K.}
\affiliation{Perimeter Institute for Theoretical Physics, 31 Caroline Street North, Waterloo, ON N2L 2Y5, Canada.}

\date{1st March 2011}

\begin{abstract}
  Randomness is an important resource for many applications, from
  gambling to secure communication.  However, guaranteeing that the
  output from a candidate random source could not have been predicted
  by an outside party is a challenging task, and many supposedly
  random sources used today provide no such guarantee.  Quantum
  solutions to this problem exist, for example a device which
  internally sends a photon through a beam-splitter and observes on
  which side it emerges, but, presently, such solutions require the
  user to trust the internal workings of the device.  Here we seek to
  go beyond this limitation by asking whether randomness can be
  generated using untrusted devices---even ones created by an
  adversarial agent---while providing a guarantee that no outside party
  (including the agent) can predict it.  Since this is easily seen to
  be impossible unless the user has an initially private random
  string, the task we investigate here is \emph{private randomness
    expansion}.

  We introduce a protocol for private randomness expansion with
  untrusted devices which is designed to take as input an initially
  private random string and produce as output a longer private random
  string.  We point out that private randomness expansion protocols
  are generally vulnerable to attacks that can render the initial
  string partially insecure, even though that string is used only
  inside a secure laboratory; our protocol is designed to remove this
  previously unconsidered vulnerability by privacy amplification.  We
  also discuss extensions of our protocol designed to generate an
  arbitrarily long random string from a finite initially private
  random string.  The security of these protocols against the most
  general attacks is left as an open question.
\end{abstract}

\maketitle

\section{Introduction}
\label{sec:intro}
Random numbers are important in a wide range of applications.  In
some, for example statistical sampling or computer simulations,
pseudo-randomness may be sufficient.  However, in others, such as
gambling or cryptography, the use of pseudo-randomness may be
detrimental|a shrewd adversary might identify and exploit any
deviation from true randomness.  Since quantum measurements are the
only physical processes we know of that appear to be intrinsically
random, it is natural to try to design quantum random number
generators.  In fact, devices which generate randomness through
quantum measurement are commercially available\footnote{For example,
  \url{www.idquantique.com}.}.  However, to be convinced that the
outputs of these devices are random and \emph{private}, i.e.\ unknown
to any third party, the user must either trust or verify that
they are built to a specified design.

It would be desirable if users could instead guarantee the privacy of
their newly generated random strings solely by tests on the outputs of
their devices.  This would eliminate the need for a complicated and
time-consuming verification that the devices are functioning
according to design and contain no accidental or deliberate security
flaws.  A protocol requiring only tests on device outputs is said to
be \emph{device-independent}.

The notion of device-independent cryptography was first introduced by
Mayers and Yao~\cite{MayersYao}, although with hindsight it could be
argued that the seed of the idea was already implicit in the Ekert key
distribution protocol~\cite{Ekert}.  Proving device-independent
security of cryptographic tasks is a challenging task.  The first
quantum key distribution protocol with proven device-independent
security was devised by Barrett et al.\ ({\sc
  BHK})~\cite{BHK}\footnote{See also Ref.~\cite{BKP} for some further
  details and discussion.}.  Although the {\sc BHK} protocol provided
a crucial proof of principle, it achieves provable general security at
the price of low efficiency.  The idea was subsequently developed,
producing more efficient protocols provably secure against restricted
classes of attack~\cite{AGM,AMP,ABGMPS,PABGMS,McKague} and then
against general
attacks~\cite{MRC,Masanes,HRW2,MPA,HR}\footnote{\label{ft:ns}These
  latter protocols have an important difference from those in the
  former set: they require that at least one of the users ensures
  additional no-signalling conditions that require multiple isolated
  regions within their laboratories.  Specifically, they are valid
  only if (for at least one of the users) each input is made to a
  separate device unable to communicate with the others.}.

Here we consider a different task, \emph{private randomness
  expansion}.  The aim is to use an initially private random string to
generate a longer one, in a way that guarantees that the longer string
is also kept private from all other parties.  In this paper, we
investigate the task of private randomness expansion within the
device-independent paradigm.

This task was first introduced in~\cite{ColbeckThesis} (the work
presented here is essentially a condensed version of Chapter~5
of~\cite{ColbeckThesis}) and has been subsequently
developed~\cite{PAMBMMOHLMM}.  In the latter work, Pironio et al.\
analyse a protocol related to the one in~\cite{ColbeckThesis} (they
use the {\sc CHSH} inequality instead of {\sc GHZ} tests) and present
a security analysis for a restricted class of attacks (ones in which
an adversary is forced to measure any quantum systems they hold prior
to performing privacy amplification).  Furthermore, they report an
experimental demonstration of their protocol.

Quantum private randomness expansion is an important cryptographic
task in its own right, but also has some features in common with
quantum key distribution, so device-independent protocols and security
proofs for this task should also shed light on analogous results for
quantum key distribution.  Conversely, any secure protocol for
device-independent key distribution that generates a secret key longer
than the amount of randomness used in the protocol could also be used
for randomness expansion by performing both sides of the
key-generation protocol in a single laboratory.  Candidate protocols
of this type have recently been proposed (see above); at present they
require a large number of isolated devices (cf.\
Footnote~\ref{ft:ns}).  Obviously, it would be preferable not to require this practically challenging
constraint, all else being equal.\footnote{Analyzing these
recent protocols and their implications for randomness expansion 
goes beyond the work reported here;
however, readers should be aware of their existence.}   

There is one significant new insight in the present work that has not
appeared previously: the protocols given in~\cite{ColbeckThesis}
and~\cite{PAMBMMOHLMM} are not secure in a composable way.  On the
contrary, there are quite plausible scenarios in which the final
private random string output by these protocols can become partly
compromised, in which case the protocol is evidently insecure.  The
protocol we present here has hence been slightly modified from the one
given in~\cite{ColbeckThesis} (see Section~\ref{sec:prot} for further
explanation).\footnote{The protocol
in~\cite{PAMBMMOHLMM} can also be modified to deal with this
security loophole, as we discuss in Footnote~\ref{tactic} below.}

Our protocol is intended to allow an honest user, Bob, to input a
sufficiently long initial private random string to devices constructed
by a potential adversary, Eve, and obtain as output a finite longer
private random string.  We also propose using this protocol within an
extended one to allow an initial private random string to generate an
arbitrarily long private random output string.  Our extended protocol
has the undesirable feature of requiring a large number of devices
(dependent on the amount of expansion required) which must
be prevented from communicating with one another.  In both protocols, the
length of initial string required depends on the tolerance for risking
successful cheating by Eve.  Neither protocol is optimized for
efficiency.

Proving security of our protocols against the most general possible
attacks remains an open question.  The aim of this work is rather to
introduce the task, to propose some candidate protocols for its
solution, and to explain some intuitions that suggest they are good
candidates to examine further.  In so doing we should stress that,
while the history of quantum cryptography shows that initially
unproven intuitions can spark major advances|indeed the subject was
originally founded on such intuitions~\cite{wiesner,bb84}|it also
shows that they should be approached critically and finally accepted
only if and when proven.

\section{Preliminaries}

\subsection{Assumptions}
We make the following assumptions:
\begin{enumerate}
\item \label{ass1} Bob's laboratory is secure.   In particular,
  secret messages cannot be sent from Eve's devices, once within his
  laboratory, to the outside world\footnote{Without this, the task is
  impossible, since the devices could simply broadcast their inputs
  and outputs.}, and Eve cannot probe his
  laboratory from outside.
\item \label{ass2} Bob can isolate any devices in his laboratory, preventing
  them from sending any signal outside an isolated region.\footnote{For example, by placing them each in
  their own sub laboratory.  Alternatively, when it is sufficient to
  prevent communication between the devices during a protocol of
  finite duration, if we assume the impossibility of faster-than-light
  signalling, Bob can isolate the devices by placing them at
  appropriately space-like separated locations during the protocol.
  (Bob could ensure such separation using trusted classical clocks and
  rulers.)}
\item \label{ass3} Bob has secure classical information
  processing devices\footnote{If Bob cannot trust any classical information
    processing device|including his own brain|then he is beyond the
    help of cryptographers.}, with secure authenticated
  classical communication between them within his laboratory.\footnote{Since Eve's devices
    can be isolated (cf.\ Assumption~\ref{ass2}), we assume that any
    authentication procedures used do not need to consume secret
    randomness; of course, if they do, this randomness should be
    included in the accounting.}
In  particular, Eve's devices are unable to spoof classical
  communications within Bob's laboratory; they can output to 
Bob's classical devices only via prescribed channels.

\item Eve is constrained by the laws of
  quantum theory.

\item \label{ass5} All communication channels and devices operate
  noiselessly\footnote{This assumption is in practice unrealistic
    and one would hope to eventually drop it.  Its main purpose is to
    keep the protocols and security proofs as simple as possible in
    the first instance.}.
\end{enumerate}

Note that we do not include the common assumption that Bob has
complete knowledge of the operation of the devices he uses to
implement the protocol.  Instead, we suppose that all quantum devices
were sourced from an untrusted party, Eve, who may construct them
using complete knowledge of Bob's protocol.  From Bob's
perspective, these devices are simply black boxes with inputs and
outputs.    

We briefly consider weakening the assumption that Eve is constrained
by quantum theory at the end of the paper.

\subsection{Non-classical correlations}\label{sec:2b}
Bob will want to perform tests on the devices supplied by Eve.  We
assume Bob's testing protocol is known publicly, and in particular is
known to Eve, but that it may involve random inputs which are not
known to Eve.  Indeed, a little thought shows that this is essential
for any unconditionally secure protocol.  Without private random
inputs, Eve knows Bob's entire protocol.  To be useful, the protocol
must have at least one valid set of outputs.  Eve can then calculate
such a set in advance and supply her devices with classical records of
these pre-calculated outputs, thus ensuring both that the devices pass
Bob's tests and that she knows in advance all the data they generate
for Bob.  Clearly, Bob cannot generate any private random data in this
scenario.

We thus assume that Bob begins with a private random string, and is
interested in generating a longer one, i.e.\ in {\it private
  randomness expansion}.  He needs to ensure that Eve cannot
pre-calculate classical data that she can supply to her devices in
order to pass his tests|otherwise she can predict all the output data
that will be generated for any given random input, and so he cannot
generate any new private randomness.  Bob must thus ensure that his
tests cannot be passed (except perhaps with a small probability) by devices
whose outputs can be described by a local hidden variable model.  To
do so, Bob needs to perform some form of \emph{Bell test}, in which
the devices are prevented from signalling to one another, either by
physical barriers or by being space-like separated, to ensure the
presence of non-classical correlations.  Secure private randomness
expansion is thus impossible without Bell tests.  

Our intuition is that, conversely, in suitable protocols, Bell tests
make private randomness expansion possible.  Roughly speaking, the
underlying idea is that states that produce non-classical correlations
possess some intrinsic randomness, uncorrelated with any other system
in the universe.  So, by verifying the presence of such correlations,
Bob can be sure that Eve's devices are using such states and hence
that he derives genuine private randomness from them.  The hypothesis
is then that, in order to pass Bob's verification with a significant
probability of success, Eve's strategy must be so close to the honest
one that she cannot gain significant information about Bob's newly
generated private randomness.

The protocols used in this work are based on the following test, which
we call a {\sc GHZ} test~\cite{GHZ}\footnote{Other tests of
  non-locality could also be used, some of which are discussed in
  Section~\ref{effic} (see also~\cite{PAMBMMOHLMM}).}.  Bob asks for
three devices, each of which has two settings (which we label $P_i$
and $Q_i$ for the $i$th device) and can output either $1$ or $-1$.  We
use $p_i$ and $q_i$ to denote the values of the outputs when inputs
$P_i$ and $Q_i$ are made.  Bob chooses one of the four triples of
settings given by $P_1P_2P_3$, $P_1Q_2Q_3$, $Q_1P_2Q_3$ and
$Q_1Q_2P_3$, obtaining a result given by the product of outputs
corresponding to the specified inputs: for example, if his inputs are
$P_1 P_2 P_3$ he obtains outcomes $p_1$, $p_2$ and $p_3$.  He demands
that the product $p_1 p_2 p_3$ is $-1$, while $p_1q_2q_3$, $q_1p_2q_3$
and $q_1q_2p_3$ are $+1$.  That these cannot be satisfied by a
classical assignment~\cite{GHZ} can be seen by considering the product
of the four quantities.  According to Bob's demands, this must be
$-1$, while the algebraic expression obtained by a classical
assignment is $p_1^2p_2^2p_3^2q_1^2q_2^2q_3^2$, which must be $+1$.
If, instead, the $\{p_i\}$ and $\{q_i\}$ are obtained from the
outcomes of measurements acting on an entangled quantum state, then
Bob's demands can always be met.  In the Appendix, the complete set of
operators and states which do this is derived.  In essence, all such
operators behave like Pauli $\sigma_x$ and $\sigma_y$ operators and
the state behaves like a {\sc GHZ} state,
$\frac{1}{\sqrt{2}}\left(\ket{000}-\ket{111}\right)$, up to local
unitaries.

\section{Private Randomness Expansion}
\subsection{Security Definitions}
In this section, we define what it means for a string to be private
and random.  
We say that $S$ is a \emph{private random string} with respect to
a system $E$ if the joint state of the string and $E$ takes the form
\begin{equation}
\label{ideal_state}
\rho^I_{SE}:=\frac{1}{|S|}\sum_s\ketbra{s}{s}\otimes\sigma_E,
\end{equation}
for some state $\sigma_E$, where the sum runs over all possible
instances, $s$, of the string $S$ and the superscript $I$ stands for
``ideal''\footnote{A note on notation: we tend to use upper case
  letters to denote random variables and lower case letters for
  particular instances of these random variables.  For random variable
  $X$, we use $|X|$ to denote the number of possible outcomes of $X$.
  Thus, $|S| = 2^n$ for a bit string $S$ of length $n$.}.  The key
properties of this state are that the $E$ system is uncorrelated to $S$,
and that the possible instances of $s$ are uniformly distributed.

In practice, it will not be possible to guarantee a state of this form.
Instead, we may have a state
\begin{equation}
\label{actual_state}
\rho^R_{SE}:=\sum_s P_S(s)\ketbra{s}{s}\otimes\rho_E^s,
\end{equation}
where the superscript $R$ stands for ``real''.  We say that the string
$S$ in this state is a \emph{$\delta$-private} random string with respect to
$E$ if there exists a $\sigma_E$ such that
$D(\rho^I_{SE},\rho^R_{SE})\leq\delta$, where
$D(\rho,\tau):=\frac{1}{2}\text{tr}|\rho-\tau|$ is the trace distance.
The trace distance is related to the optimal probability of guessing
which of two states one has.  Its operational significance is that if
$D(\rho,\tau)\leq\delta$, then no physical procedure allows one to
distinguish between $\rho$ and $\tau$ with success probability greater
than $\delta$.  Moreover, since the trace distance is non-increasing
under quantum operations~\cite{Ruskai}, this condition must persist
when the string is used in any application.

In private randomness expansion, typically the raw outputs of the devices
are not $\delta$-private for a sufficiently small $\delta$, and
require \emph{privacy amplification} in order to reduce $\delta$ to an
acceptable level for security.  This is described in detail in the
next subsection.

It is impossible to devise a finite device-independent cryptographic
protocol that guarantees non-trivial security for any task with {\it
  complete} certainty.  Eve can always follow the strategy of guessing
the random input string and supplying appropriate pre-computed outputs:
this has a nonzero probability of success.  In particular, it is
impossible to construct a private randomness expansion protocol that
\emph{guarantees} that the final string is $\delta$-private (for small
$\delta$) against an
arbitrary attack by Eve.  Our security criterion thus involves two
parameters.  We demand that for any strategy chosen by Eve, the a
priori probability that the protocol does not abort and the final
string is not $\delta$-private is at most $\zeta$, where $\delta$ and
$\zeta$ are suitably small.  We say that a protocol with this property
is a {\it $\zeta$-secure} protocol that generates a $\delta$-private
string\footnote{A protocol that never aborts is thus $\zeta$-secure
  if the a priori probability that the final string is
  $\delta$-private is at most $\zeta$.}.  Since $\delta$ and $\zeta$
are small, Eve has only a small probability of learning a significant
amount of information about the final string without causing the
protocol to abort.

\subsection{Privacy Amplification}
\label{priv_amp}
Privacy amplification takes an initial string, $X'$, about which a
potential adversary has partial knowledge, $E$, and compresses it to a
shorter string, $S$, which is approximately uniformly distributed and
independent of the adversary's knowledge.  This typically requires
some additional randomness, $R$, to select the function, $f$, from
some set of functions, $\cF$, used for the compression.  The idea is
that by choosing the set $\cF$ appropriately, the final string, $S$,
is very close to being private and random (according to the definition
given in the previous section).  Furthermore, we require that the
string $S$ is very close to independent of the randomness, $R$, used to choose the
function.

Privacy amplification was first studied in the case where the
adversary's knowledge, $E$, is classical (see for
example~\cite{BBR,ILL,BBCM}) and was later extended to the case of
quantum knowledge~\cite{KMR,RennerKoenig,Renner}.  In the latter
works, it was shown that the length of extractable private random
string can be characterized in terms of the smooth conditional
min-entropy of the initial string, $X'$, given the knowledge, $E$, the
quantum version of which was first introduced in~\cite{Renner}.

The smooth min-entropy can be defined not only for strings, $X'$, but
for any quantum state on a system, $B$.  We first define the
non-smooth min-entropy of $B$ given $E$ for a state $\rho_{BE}$:
\begin{equation*}
H_{\min}(B|E)_{\rho}:=\max_{\sigma_E}\sup\{\lambda\in\mathbb{R}: 2^{-\lambda}\id\ot\sigma_E\geq\rho_{BE}\},
\end{equation*}
where the maximization is over normalized density operators,
$\sigma_E$.  The $\epsilon$-smooth conditional min-entropy of $B$
given $E$ is then defined by
\begin{equation}
H_{\min}^{\epsilon}(B|E)_{\rho}:=\max_{\bar{\rho}_{BE}}H_{\min}(B|E)_{\bar{\rho}},
\end{equation}
where the maximization is over a set of positive (and potentially
sub-normalized) operators $\epsilon$-close to $\rho$ with respect to
some distance measure\footnote{Various distance measures have been
  used in the past.  One popular approach is to use the \emph{purified
    distance}, which is the minimum trace distance over purifications
  of the involved states (see~\cite{TCR2} for further details).}.

We consider the use of two-universal hash functions for privacy
amplification which are defined as follows~\cite{CW,WC}:
\begin{definition}
  A set of functions, $\cF$ from $X'$ to $S$ is \emph{two-universal}
  if when $f_r\in \cF$ is picked using a uniform random variable $R$,
  for any distinct instances, $x'_1$ and $x'_2$, of $X'$, the
  probability that they give the same values of $S$ is at most
  $\frac{1}{|S|}$, i.e.,
  $\frac{|\{r:f_r(x'_1)=f_r(x'_2)\}|}{|R|}\leq\frac{1}{|S|}$.
\end{definition}

We remark that other appropriate functions could be used instead|we
would like a class of functions with the fewest members.  In
particular, privacy amplification schemes based on Trevisan's
extractor~\cite{trevisan} have recently been shown to be secure even
when the information $E$ is quantum, and in general require a shorter
seed than schemes using two-universal hash functions~\cite{DPVR}.  To
simplify the discussion, given that we do not analyse optimality or
security against general attacks, we focus on two-universal hashing
here.

Including the classical spaces used to define the string, $X'$, and
the random string used to choose the hash function, $R$, the state we
have prior to privacy amplification has the form
\begin{equation}\label{eq:postmeas}
\rho_{X'ER}=\sum_{r,x'}P_R(r)P_{X'}(x')\ketbra{x'}{x'}\otimes\rho_E^{x'}\otimes\ketbra{r}{r},
\end{equation}
where $P_R(r)=\frac{1}{|R|}$.  After applying the hash function
$f_r\in \cF$, the state takes the form
\begin{equation}\label{rhoser}
\rho_{SER}=\sum_{r,s}P_R(r)P_S(s)\ketbra{s}{s}\otimes\rho_E^{s,r}\otimes\ketbra{r}{r},
\end{equation}
where $P_S(s)\rho_E^{s,r}=\sum_{x': f_r(x')=s}P_{X'}(x')\rho_E^{x'}$.
Ideally, the state of the system in $\mathcal{H}_S$ would look uniform
from Eve's point of view, even if she were to learn $R$ (functions for
which this property holds are sometimes called \emph{strong
  extractors}).  The variation from this ideal can be expressed in
terms of the trace distance between the state and an ideal,
\begin{equation}
D(\rho_{SER},\tau_S\otimes\sigma_{ER}) \, ,
\end{equation}
where $\tau_S$ is the
maximally mixed state in $\mathcal{H}_S$ and $\sigma_{ER}$ is an
arbitrary state.  This distance is bounded in the following
theorem~\cite{Renner}.
\begin{theorem}
\label{Ren1}
If $f_r$ is chosen from a two-universal set of hash
functions, $\cF$, using a uniform random string, $R$, which is
uncorrelated with $S$ and $E$, and is used to map $X'$ to $S$ as
described above, then for $|S|=2^{t}$ and $\epsilon\geq 0$, we have
\begin{equation}
\label{sm_min_ent}
\min_{\sigma_{ER}}D(\rho_{SER},\tau_S\otimes\sigma_{ER})\leq\epsilon+\frac{1}{2}2^{-\frac{1}{2}\left(H_{\min}^{\epsilon}(X'|E)-t\right)}.
\end{equation}
\end{theorem}

Hence, if Bob chooses $t=H_{\min}^{\epsilon}(X'|E)-\ell$, for some
$\ell\geq 0$, he can use a random string, $R$, to compress his string,
$X'$, which is partly correlated with a quantum system held by Eve, to
a $\delta$-private string, $S$, for some
$\delta\leq\epsilon+\frac{1}{2}2^{-\frac{\ell}{2}}$.

We remark that privacy amplification is usually discussed in a three
party scenario, in which Alice and Bob seek to generate a shared
random string on which Eve's information is negligible.  Alice and Bob
are required to communicate during the amplification stage, and thus
leak information (in our case the random string $R$) about the
amplification process to Eve.  Private randomness expansion, on the
other hand, is a task involving only Bob, who aims to generate data
secret from Eve.  No information need be leaked in amplification since
there is no second honest party needing to perform the same procedure.
The random string $R$ hence remains private with respect to
Eve\footnote{In principle, Eve can gain a little information about
  $R$ if and when she learns $S$, but only within the tight privacy
  bounds implied by~\eqref{rhoser} and~\eqref{sm_min_ent}.}.

\section{Protocols}\label{sec:prot}
We begin this section by giving a protocol which is designed to allow
a private random string to be expanded by a finite amount.  Before
performing the protocol, Bob asks Eve for three devices, each of which
has two settings (inputs), ($P_i$ and $Q_i$ for the $i$th device) and
can make two possible outputs, $+1$ or $-1$.  Bob asks that whenever
these devices are used to measure one of the four {\sc GHZ} quantities
($P_1P_2P_3$, $P_1Q_2Q_3$, $Q_1P_2Q_3$ and $Q_1Q_2P_3$), they return
outcomes with the properties specified in Section~\ref{sec:intro}
(i.e., whose products are $-1$, $+1$, $+1$ and $+1$
respectively)\footnote{In practice, Bob might ask for devices that
  measure either $\sigma_x$ or $\sigma_z$, and for a further device
  that creates {\sc GHZ} states.  However, he will not be able to
  distinguish such devices from another set satisfying the test but
  using a different set of states and operators.  We have kept the
  description in terms of things he can verify.}.  Furthermore, he asks
that these devices can satisfy these conditions without communicating.
We call these three devices taken together a device triple.  Bob uses
his device triple in the following protocol.

\subsection*{Protocol~1}
This protocol depends on parameters $\zeta\geq 0$ and
$\delta\geq 0$ and can be applied to an initial private string
$X$\footnote{For any finite length of initial string, there will be 
minimum values of $\zeta$ and $\delta$ below which the protocol never
increases the length of the string.}.
Although we express it for the case of {\sc GHZ} tests and
two-universal hashing, it is easily adapted to other Bell tests or
privacy amplification functions.

\begin{enumerate}
\item \label{first} Bob sets up the device triple such that the
  devices cannot communicate with one another (cf.\
  Assumption~\ref{ass2}), nor send any information outside Bob's
  laboratory (cf.\ Assumption~\ref{ass1}).
\item \label{second} Bob divides his string $X$ into two strings $X_1$
  and $R$ (the relative lengths of these strings depends on the choice
  of function used for privacy amplification in Step~\ref{PA}).
\item \label{pst1} Bob uses two bits of $X_1$ to choose one of the four
  tests which he performs, ensuring that each device learns only its
  input (and not the whole string $X_1$).  He brings all the output bits
  together.
\item \label{pst3} If he
  receives the wrong product of outputs, he aborts\footnote{In a more
    general protocol tolerating noise, Bob need not abort.  Instead,
    he would collect statistics on when the devices generate outcomes
    with the wrong product, and use these to bound the min-entropy in
    Step~\ref{PA}.}, otherwise he turns his output into two bits
  using an appropriate assignment (for each test there are four
  possible valid output combinations).  In this way, Bob builds a
  random string $\tilde{X}$.
\item \label{pst4} Bob repeats steps~\ref{pst1} and~\ref{pst3} until
  he has exhausted $X_1$.
\item \label{PA} Bob concatenates $X_1$ and $\tilde{X}$ to form a new
  string $X'$.  He computes a suitable value
  $\gamma:=\gamma(|X_1|,\zeta,\delta)$, where $|X_1|$ is the number of
  possible input strings, $\delta$ is the desired privacy parameter
  for the output string, and $\zeta$ is the a priori risk he will
  tolerate that the output string is not $\delta$-private\footnote{In
    noise-tolerant protocols, $\gamma:=\gamma(|X_1|,T,\zeta,\delta)$,
    where $T$ is the number of tests passed.  In either the noiseless or
    the noise-tolerant case, finding an explicit form of the function
    $\gamma$ that provably has the desired properties remains an open
    problem: see the comments later in this section.  Note also that
    $\gamma$ may be zero if any of the parameters are too small.}.  He
  performs privacy amplification on $X'$ to form a string of length
  $\log|S|=\gamma$ bits
  which, with a priori probability greater than $(1 - \zeta)$, is
  $\delta$-private.  In the case of two-universal hashing, $R$ and
  $X'$ have equal length~\cite{RennerKoenig,Renner}\footnote{More
    recently, it has been shown that a shorter $R$ roughly equal to
    the size of the output, $S$, can be used~\cite{TSSR}.  In the
    context of the present protocol, unless there are significant
    levels of cheating or noise, we do not expect that this will have
    a significant effect on the rate.} and so Bob should partition
  $X=(X_1,R)$ such that $2\log_2|X_1|=\log|R|$.
\item The protocol's output is the concatenated string $(S,R)$.
\end{enumerate}

\begin{figure}

\includegraphics[width=0.5\textwidth]{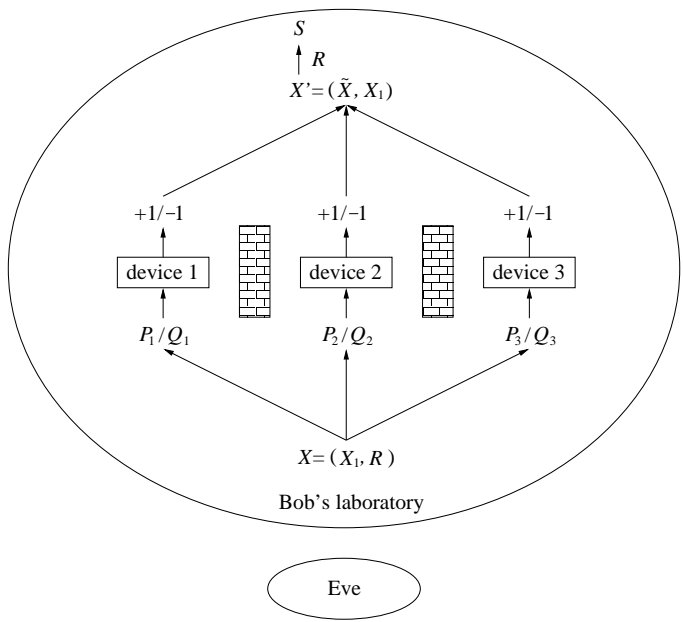}
\caption{Diagram of the steps in Protocol~1.  Together devices 1--3
  form a device triple.  They are prevented from communicating with
  one another (depicted by the walls) or to the outside world (Eve).
  Each device learns only the set of inputs it is supplied with.}
\label{fig:Rand_exp1}
\end{figure}

The entire setup is shown in Figure~\ref{fig:Rand_exp1}.

\mbox{}

If Eve is constrained by quantum theory, then the only way she can be
certain to pass all of Bob's tests is if the joint state shared by the
devices is pure and generates unbiased outcomes (see the Appendix).
This strategy gives no information to Eve.  Moreover, in this case,
two bits of $X_1$ generate two new bits of randomness each time the
loop is performed.  This is an attractive feature of the GHZ-based
protocol: the same operations are used both to test security and to
generate new random bits.  Furthermore, if Bob trusts Eve, he can
forego the privacy amplification step and the shorter protocol is very
efficient, doubling the length of the random key.

That said, of course, the aim is to protect Bob against a potentially
dishonest Eve who can prepare the devices to include any quantum
systems, which may be entangled with one another and also with an
ancillary system kept under Eve's control.  She may also prepare the
devices with any quantum program to produce outputs from
inputs\footnote{In principle, Eve may also design the devices so as
  to attempt to send any quantum signals to one another, according to
  any algorithm of her choice.  However this is pointless if, as in
  our protocol, Bob prevents such signalling.}.

It remains an open problem to find a function $\gamma(|X_1|,
\zeta,\delta)$ (or $\gamma(|X_1|, T, \zeta,\delta)$ in the
noise-tolerant case) such that the protocol is $\zeta$-secure.  Since
Eve is constrained by quantum theory, the joint quantum state of the
system Bob uses to store $X'$ and Eve's systems has the form
$\sum_{x'}P_{X'}(x')\ketbra{x'}{x'}\ot\rho_E^{x'}$ prior to privacy
amplification.  Here, the information, $E$, should include any
additional information about the protocol that Eve might possibly
infer from data that Bob is not required by the protocol to keep
private (and in realistic applications may not necessarily be able to
keep private): for example, the length of the final private random
string, whether the protocol aborted, or how many rounds were
performed\footnote{\label{ft:priv}One can imagine scenarios in which
  no such information is in fact ever leaked to Eve.  However, for
  unconditional and composable security, Bob's final string should
  remain private and random even if such information becomes known to
  Eve. See the discussion at the end of this section for some
  realistic scenarios in which this concern applies.}.

We would then like a statement which says that, for any $\zeta > 0$,
there exists some calculable $\ell \geq 0$ such that, for any strategy
used by Eve, we have $p \leq \zeta $, where $p$ is the probability 
(averaged over all possible initial
random strings $X$) that $(i)$ the protocol does not abort and 
$(ii)$ the
min-entropy fails to satisfy
$H_{\min}^{\epsilon}(X'|E)\geq\gamma-\ell$.  
Conversely, if the min-entropy satisfies this bound, it follows that
the string $S$ (of length $\gamma$) formed by hashing $X'$ is
$\delta$-private for
$\delta=(\epsilon+\frac{1}{2}2^{-\frac{\ell}{2}})$ (see
Theorem~\ref{Ren1})\footnote{Since it is only quantum devices that
  are supplied by Eve, and hashing is a classical procedure, there is
  no security issue associated with this step.}.

Intuition suggests|as a working hypothesis awaiting full
analysis|that, in order to ensure a reasonable probability of the
protocol not aborting, Eve cannot deviate much from her honest
strategy, and so the string produced by a successful run of the
protocol almost certainly satisfies $H_{\min}^{\epsilon}(X'|E)\lesssim
2\log_2|X_1|$ for some suitably small $\epsilon$.  The length of the
final output string would then be
$\log_2|R|+t\approx\frac{4}{3}\log_2|X|-\ell$, i.e.\ the protocol
would increase the length of the string by a factor close to
$\frac{4}{3}$. (For a long enough initial string, a given level of
security can be achieved with $\ell\ll\log_2|X|$.)

It is important to note that the string generated by our protocol,
although private with respect to Eve, is not private with respect to
the devices, which could be programmed to remember their output bits.
The generated random string thus cannot be treated as defining an
effectively independent new input string for the same devices.
Furthermore, it is important that Bob prevents the devices from
sending signals outside his laboratory until the private randomness is
no longer required.

\subsection*{Attacks on the initially private string}
In earlier work on randomness
expansion~\cite{ColbeckThesis,PAMBMMOHLMM} it was argued that there is
no need to include $X_1$ in the string undergoing privacy
amplification, the argument being that since $X_1$ is only used to do
operations within Bob's laboratory, it always remains secure against
the outside and hence can be included in the final private string
without any processing.  However, we argue here that there are
reasonable scenarios in which, on the contrary, Eve {\it could} learn
part of $X_1$ despite the security of Bob's laboratory.  Protocols in
which $X_1$ is left unprocessed in the final string do not have
universally composable security, and indeed are evidently insecure in
some reasonable applications.

We illustrate this point by giving a particular strategy
which gives Eve a significant probability of learning some bits of
$X_1$.  Suppose Eve programs the devices such that the protocol aborts
unless a set of $m$ specified bits of $X_1$ take specific
values\footnote{Eve can do this by fixing pre-specified outputs whose
  product is $-1$ for these bits, so that they are valid outputs for
  input $P_1P_2P_3$ but not for any other input.  Alternatively, she
  can pre-specify outputs that are invalid: for example, the devices
  could be programmed to output $2$ (or to fail to make an output) for
  any input except $P_1Q_2Q_3$.}.  For small $m$, Eve can ensure the
devices behave in this way while keeping the probability of the
protocol not aborting significantly above zero.  If the string $X_1$
remained a truly a private random string, this property should persist
if Bob announces whether the protocol aborted or not (see also
Footnote~\ref{ft:priv}).  However, if Eve uses this strategy, the knowledge that the protocol did not abort would convey
to Eve the values of the $m$ specified bits of $X_1$\footnote{Eve also
  has more general attacks of this type that allow her to learn some
  information about some bits: for instance, she can pre-specify some
  outputs whose product is $1$, which will be valid for inputs
  $P_1Q_2Q_3$, $Q_1P_2Q_3$ and $Q_1Q_2P_3$ but will cause an abort (in
  the noiseless case) if the input is $P_1P_2P_3$.}.

To make this point more concrete, imagine that Eve knows that Bob's
casino relies on purportedly private random bits that are output from
this protocol for tonight's operations.  If the protocol aborts, Eve
knows the casino will not open tonight.  However, there is a
significant chance that the protocol will not abort, and the casino
will open tonight.  Moreover, Eve knows that, if the casino does open,
it will continue to run until the purportedly private random bit
string that Bob has generated is exhausted.  Eve can thus gain
information about some bits conditioned on the casino opening, or
staying open, and profitably bet on the relevant bits.  Clearly, this
is not consistent with a sensible definition of private randomness.
Note that the protocol in~\cite{PAMBMMOHLMM} is equally vulnerable to
attacks of this kind\footnote{Strictly speaking, as presented, the
  protocol in~\cite{PAMBMMOHLMM} never aborts, although it may fail to
  increase the length of the initially private random seed string.
  However, it is vulnerable to an attack in which Eve uses the length
  of the final generated key to infer information about the
  purportedly still private random seed.   A fix for this is discussed
in the following footnote.}.

In our protocol, the idea is to avoid this problem by
performing privacy amplification on $X_1$ as well as on $\tilde{X}$.
Another possible strategy is to look for protocols that are efficient
enough in generating new randomness that even if $X_1$ is discarded
prior to privacy amplification, the final private random string is
longer than the original.  The original random string can then simply be
discarded after use.  Protocols based on higher dimensional generalizations of the {\sc
  GHZ} test appear to be good candidates of this type (see
the next section).  Other classes of candidates are protocols in which
the inputs to the devices correspond to tests only on a relatively
small (randomly chosen) subset of the rounds, with deterministic
inputs used for the remainder, or to protocols in which the inputs are
chosen with a non-uniform distribution that requires a relatively
small amount of randomness to sample from.\footnote{\label{tactic}This tactic is
used in the protocol in~\cite{PAMBMMOHLMM}.  Although, as presented,
the protocol in~\cite{PAMBMMOHLMM} is vulnerable to the attack
discussed here, this
could be fixed by simply discarding the relevant part of the initial string.
Doing so makes the potential insecurity of the initial string
irrelevant, but of course may significantly affect the accounting in practical 
implementations.  For example, if applied to the reported experiment in~\cite{PAMBMMOHLMM}, 
it would mean the final private random string produced is actually
shorter than the initial private random string.}

\subsection*{Iteration via Isolation}

The protocol we have presented aims to expand a finite initial random string
by a finite additional amount.  However, it is natural to ask whether
indefinite expansion of a finite random string is possible.  We now present an extended protocol
which suggests that this may, in principle, be
achievable\footnote{Again, we stress that rigorous analysis remains a
  task for the future.}.  However, this protocol has the disadvantage
that it requires a large number of additional devices.  In the
extended protocol, Bob asks Eve for $N$ device triples and arranges
them such that no two can communicate, e.g.\ by placing them in their
own sub-laboratories.  He then performs Protocol~1 using the first
device triple, generating a new string $(S,R)$.  This is then used to
perform Protocol~1 on the second device triple and so on\footnote{A
  word of caution: at present Theorem~\ref{Ren1} is only proven to
  hold in the case that the randomness used to choose the function is
  perfectly uniform and perfectly uncorrelated with $S$ and $E$.}.
To get started, this extended protocol requires that the initial
private random string is sufficiently long that it can be securely
expanded.

\section{Tests Based on other Correlations}
\label{effic}
In this section we discuss some alternative ways of constraining Eve
rather than demanding that her outputs satisfy {\sc GHZ} tests, with a
view to improving the rate (i.e., the length of private random string
generated by a given length of initial private random string) while
keeping the number of devices required relatively small.  One
promising family of correlations come from direct generalizations of
the {\sc GHZ} correlations to more parties, as conceived by Pagonis,
Redhead and Clifton ({\sc PRC})~\cite{PRC}.  Their family of tests is
such that in the $k$th version of this test, $4k-1$ devices are
required to measure one of $4k$ quantities (i.e., $\log_2(4k)$ bits of
randomness are required per test), while generating $4k-2$ bits of
randomness. (The case $k=1$ corresponds exactly to the {\sc GHZ}
test.)

For example, in the case $k=2$, Bob asks for seven of the two-input,
two-output devices discussed previously and considers measuring one of
the the eight combinations
\begin{flushleft}
$P_1Q_2Q_3Q_4Q_5Q_6Q_7,\quad Q_1P_2Q_3Q_4Q_5Q_6Q_7$,
$Q_1Q_2P_3Q_4Q_5Q_6Q_7,\quad Q_1Q_2Q_3P_4Q_5Q_6Q_7$,
$Q_1Q_2Q_3Q_4P_5Q_6Q_7,\quad Q_1Q_2Q_3Q_4Q_5P_6Q_7$,
$Q_1Q_2Q_3Q_4Q_5Q_6P_7,\quad P_1P_2P_3P_4P_5P_6P_7$.
\end{flushleft}
He demands that the products of the outputs for the first seven
combinations are always $+1$ and for the last combination, the product
of the outputs should be $-1$.  This can be achieved using {\sc PRC}'s
$7$-party generalization of the {\sc GHZ} state.  For this test, 3
bits of randomness are required to choose amongst the eight settings,
while in a successful implementation of the test on this state, 6 bits
of randomness are generated by the output.  If Bob trusts Eve, the
private random string is tripled in length and for larger $k$, the
expansion is even more dramatic.

Such tests thus look like very good candidates for private randomness
expansion.   However, of course, we still need to introduce privacy
amplification to protect Bob against a dishonest Eve.
Using $\log_2|X_1|$ bits of private randomness we can
perform $\frac{\log_2|X_1|}{\log_2(4k)}$ tests, generating
(approximately) $\frac{\log_2|X_1|}{\log_2(4k)}(4k-2)$ new bits.  If
one uses two-universal hashing for privacy amplification, then in
order to choose the hash function, we require
$\log_2|R|=\frac{\log_2|X_1|}{\log_2(4k)}(4k-2)+\log_2|X_1|$ bits.  We
also have $\log_2|X|=\log_2|X_1|+\log_2|R|$, hence the length of final
string is (approximately) $\frac{2k-1}{\log_2(4k)+2k-1}\log_2|X|$ bits
longer than the original.  Hence, in the limit of large $k$, we would
intuitively expect the analogous protocols to roughly double the
length of private random string.  Furthermore, if a function requiring
shorter $R$ were used for privacy amplification (e.g.\ that
of~\cite{DPVR}), the rate increase with $k$ is potentially greater. 

Even if such tests do improve the rate of random string expansion,
though, there are trade-offs.  Firstly, more devices are needed and,
secondly, it seems likely that a longer initial private string is
required in order to achieve a given level of security.  The intuition
behind this second statement comes from considering a classical
attack.  For a {\sc GHZ} test, a classical attack can escape detection
with probability $\frac{3}{4}$ per test, while in the $k$th
generalization, this increases to $\frac{4k-1}{4k}$.

One could also use a test based on the {\sc CHSH}
correlations~\cite{CHSH}, as considered in~\cite{PAMBMMOHLMM}.
CHSH-based protocols do not have the convenient cheat-detection
property that protocols based on {\sc GHZ} correlations (and their
generalizations) possess in the noiseless case: for tests based on
{\sc CHSH} correlations one can never be {\it completely} certain that
cheating has been identified.  Nor do they have the appealing property
that correlation tests can be directly used as new random bits in the
case where Eve is honest.  These features of GHZ-based protocols have,
at least, some pedagogical value, allowing as they do a simple
explanation of the basic idea of random string expansion.  However, as
far as we are aware, it is generally an open question to identify
which class of Bell violation allows the most efficient private
randomness expansion protocols for a given set of parameters $|X|$,
$\zeta$, $\delta$ and for a given costing of the various cryptographic
and physical resources involved.

\section{Discussion}
Private randomness expansion may be a useful primitive on which to
base other protocols in the untrusted device scenario.  More
fundamentally, we can think of nature as our untrusted adversary which
provides devices.  One could then argue that our protocols strengthen
the belief that nature genuinely generates
randomness\footnote{
  Of course, this assumes both that nature is constrained by the
  no-signalling principle and that we can generate some initial
  randomness uncorrelated with nature's subsequent behaviour: it is
  impossible to rule out cosmic
  conspiracy.
}.

The untrusted devices scenario is a realistic one, and seems likely to
become important if quantum computers or quantum cryptosystems become
widespread.  Ordinary users will not want to construct their own
hardware and will instead turn to suppliers, just as users of
classical computers and encryption software do today.  The protocols
in this paper are designed with the ultimate aim of offering such
users a virtual guarantee that the devices supplied are behaving in
such a way that their outputs are private and random.

Finally, we note the possibility of the given protocols being secure
even against an adversary who is not bound by quantum theory.  As {\sc
  BHK} first
showed (\cite{BHK}; see also~\cite{MRC,Masanes,HRW2}), quantum key
distribution protocols can be provably secure even against such an
adversary, provided certain signalling constraints can be guaranteed.
In the case of our private randomness expansion protocol, the
post-quantum adversary is analogously constrained: we assume that Bob
can ensure there is no signalling between any of the devices held
separately in his laboratory, nor between any of them and Eve.  It is
a further open problem to provide a security proof in this scenario.

We expect that, if additional private randomness can be securely
generated by our protocol in this post-quantum scenario, it will be at
a lower rate than in the quantum case, since Eve has more general
attacks available.

For instance, Eve can exploit the power of so-called non-local ({\sc
  NL}) boxes|hypothetical devices that maximally violate the {\sc
  CHSH} inequality.  In the notation introduced in
Section~\ref{sec:2b}, the device's outputs satisfy $p_1p_2=-1$ and
$p_1q_2=p_2q_1=q_1q_2=1$~\cite{Cirelson93,PopescuRohrlich}.  By using
{\sc NL} boxes, Eve can always know the output of one of the devices.
For example, if she sets the third device to output $1$ and the first
two to obey the {\sc NL} box conditions given above, she will always
pass a {\sc GHZ} test.  It is therefore clear that at most one bit of
private randomness would result from each test (rather than close to
two bits if Eve uses a quantum strategy).

\acknowledgments We thank Graeme Mitchison, Renato Renner and Marco
Tomamichel for helpful discussions and Lluis Masanes and Stefano Pironio for helpful
comments on an earlier draft.  This work was partially supported
by an FQXi mini-grant and by Perimeter Institute for Theoretical
Physics. Research at Perimeter Institute is supported by the
Government of Canada through Industry Canada and by the Province of
Ontario through the Ministry of Research and Innovation.

\appendix
\section*{Appendix}

The technique that we follow here is based on that used to find the
complete set of states and measurements producing maximal violation of
the {\sc CHSH} inequality~\cite{PopescuRohrlich}.

We seek the complete set of tripartite states (in finite dimensional
Hilbert spaces), and two-setting measurement devices that output
either $1$ or $-1$, such that, denoting the observables measured by
device $i$ by $\hP_i$ and $\hQ_i$, we have
\begin{eqnarray}
\label{rel1}\hP_1\otimes \hP_2\otimes \hP_3\ket{\Psi}&=&-\ket{\Psi}\\
\label{rel2}\hQ_1\otimes \hQ_2\otimes \hP_3\ket{\Psi}&=&\ket{\Psi}\\
\hQ_1\otimes \hP_2\otimes \hQ_3\ket{\Psi}&=&\ket{\Psi}\\
\label{rel4}\hP_1\otimes \hQ_2\otimes \hQ_3\ket{\Psi}&=&\ket{\Psi} \, . 
\end{eqnarray}
Here $\ket{\Psi}$ is the tripartite state.  (We consider the case of
pure states since the mixed state case follows immediately from it.)
We then have
\begin{eqnarray}
\label{ineq2}F\ket{\Psi}\equiv\frac{1}{4}\left(\hP_1\otimes \hQ_2\otimes \hQ_3+\hQ_1\otimes
\hP_2\otimes \hQ_3\,+\right.\\
\hQ_1\otimes \hQ_2\otimes \hP_3-
\nonumber
\left.\hP_1\otimes \hP_2\otimes
\hP_3\right)\ket{\Psi}&=&\ket{\Psi}.
\end{eqnarray}
$\ket{\Psi}$ is thus an eigenstate of $F$ with eigenvalue 1, so that
$F^2\ket{\Psi}=\ket{\Psi}$.  This is equivalent to
\begin{eqnarray}
\nonumber(i[\hP_1,\hQ_1]\otimes
i[\hP_2,\hQ_2]\otimes\openone+i[\hP_1,\hQ_1]\otimes\openone\otimes
i[\hP_3,\hQ_3]\,+\\\nonumber\openone\otimes i[\hP_2,\hQ_2]\otimes
i[\hP_3,\hQ_3])\ket{\Psi}=12\ket{\Psi},
\end{eqnarray}
where $[\hP,\hQ]:=\hP\hQ-\hQ\hP$ is the commutator of $\hP$ and $\hQ$.

The maximum eigenvalue of $i[\hP_1,\hQ_1]$ is 2, hence
\begin{equation}
i[\hP_1,\hQ_1]\otimes i[\hP_2,\hQ_2]\otimes\openone\ket{\Psi}=4\ket{\Psi}
\end{equation}
and similar relations for the other permutations.  We hence have
$i[\hP_1,\hQ_1]\otimes\openone\otimes\openone\ket{\Psi}=2\ket{\Psi}$ from
which it follows that
$\bra{\Psi}\left(\left\{\hP_1,\hQ_1\right\}\otimes\openone\otimes\openone\right)^2\ket{\Psi}=0$,
and hence that
$\left(\left\{\hP_1,\hQ_1\right\}\otimes\openone\otimes\openone\right)\ket{\Psi}=0$,
where $\{\hP,\hQ\}:=\hP\hQ+\hQ\hP$ is the anti-commutator of $\hP$ and $\hQ$, and we
use that $\hP_i$ and $\hQ_i$ have outcomes $\pm 1$ and hence satisfy
$\hP_i^2\ket{\psi}=\ket{\psi}$ and $\hQ_i^2\ket{\psi}=\ket{\psi}$.

Consider the following Schmidt decomposition:
$\ket{\Psi}=\sum_{i=1}^n\lambda_i\ket{i_1}\ket{i_{23}}$, where
$\lambda_i\geq 0$ $\forall$ $i$, and $n$ is the dimensionality of the
first system.  Then, if $\lambda_i\neq 0$ $\forall$ $i$, the
$\{\ket{i_1}\}$ are $n$ eigenstates of $\left\{\hP_1,\hQ_1\right\}$, each
having eigenvalue 0.  Since there are only $n$ eigenstates, we must
have $\left\{\hP_1,\hQ_1\right\}=0$.

If some of the $\lambda_i$ are zero, then we can define a projector
onto the non-zero subspace.  Call this $\Pi_1$, and define $\hp_1=\Pi_1
\hP_1\Pi_1$ and $\hq_1=\Pi_1 \hQ_1\Pi_1$.  Similarly, define projectors
$\Pi_2$ and $\Pi_3$, and hence operators $\hp_2,$ $\hq_2$ and $\hp_3$, $\hq_3$
by taking the Schmidt decomposition for systems (1,3) and 2, and (1,2)
and 3, respectively.  It is then clear that
\begin{eqnarray*}
\frac{1}{4}(\hp_1\otimes \hq_2\otimes \hq_3+\hq_1\otimes \hp_2\otimes
\hq_3+\hq_1\otimes \hq_2\otimes \hp_3-\\\hp_1\otimes \hp_2\otimes
\hp_3)\ket{\Psi}=\ket{\Psi}
\end{eqnarray*}
holds for the projected operators, and hence, these satisfy
$\{\hp_i,\hq_i\}=0$ for $i=1,2,3$.

The relations, $\hp_i^2=\openone$, $\hq_i^2=\openone$, $\{\hp_i,\hq_i\}=0$
then apply for the Hilbert space restricted by $\{\Pi_i\}$.  These
imply that $\hp_i$, $\hq_i$ and $\frac{i}{2}[\hq_i,\hp_i]$ transform like the
generators of SU(2).  The operators may form a reducible
representation, in which case we can construct a block diagonal matrix
with irreducible representations on the diagonal.  The anti-commutator
property means that only the two-dimensional representation can
appear, hence we can always pick a basis such that
$\hp_i=\openone_{d_i}\otimes\sigma_x{}_i$ and
$\hq_i=\openone_{d_i}\otimes\sigma_y{}_i$ for some dimension, $d_i$, of
identity matrix.  Our state then needs to satisfy
$\openone_{d_1}\otimes\sigma_x{}_1\otimes\openone_{d_2}\otimes\sigma_x{}_2\otimes\openone_{d_3}\otimes\sigma_x{}_3\ket{\Psi}=-\ket{\Psi}$,
and similar relations for the other combinations analogous to
(\ref{rel2}--\ref{rel4}).  By an appropriate swap operation, this
becomes
\begin{equation*}
\openone_{d_1d_2d_3}\otimes\sigma_x{}_1\otimes\sigma_x{}_2\otimes
\sigma_x{}_3\ket{\Psi}=-\ket{\Psi},
\end{equation*}
etc., which makes it clear that the system can be divided into
subspaces, each of which must satisfy the {\sc GHZ} relation (\ref{ineq2}).
In an appropriate basis, we can write
\begin{equation*}
\ket{\Psi}=\left(\begin{array}{c}a_1\ket{\psi_{\text{GHZ}}}\\a_2\ket{\psi_{\text{GHZ}}}\\\vdots\end{array}\right),
\end{equation*}
where
$\ket{\psi_{\text{GHZ}}}=\frac{1}{\sqrt{2}}(\ket{000}-\ket{111})$, and
the complex coefficients $\{a_j\}$ simply weight each subspace and
satisfy $\sum_j |a_j|^2=1$.  (Note that
$\ket{\psi_j}=\ket{\psi_{\text{GHZ}}}$ is the only solution to
$(\sigma_x{}_1\otimes \sigma_y{}_2\otimes
\sigma_y{}_3+\sigma_y{}_1\otimes \sigma_x{}_2\otimes
\sigma_y{}_3+\sigma_y{}_1\otimes \sigma_y{}_2\otimes
\sigma_x{}_3-\sigma_x{}_1\otimes \sigma_x{}_2\otimes
\sigma_x{}_3)\ket{\psi_j}=4\ket{\psi_j}$, up to global phase.)

We have hence obtained the complete set of states and operators
satisfying (\ref{rel1}--\ref{rel4}), up to local unitaries.

\end{document}